# Analysing the coverage of the University of Bologna's bibliographic and citation metadata in OpenCitations collections

Erica Andreose[1] [orcid:0009-0003-7124-9639], Salvatore Di Marzo[1] [orcid:0009-0006-0853-1772], Ivan Heibi[2,3] [orcid:0000-0001-5366-5194], Silvio Peroni[2,3] [orcid:0000-0003-0530-4305], Leonardo Zilli[1] [orcid:0009-0007-4127-4875]

[1] Digital Humanities and Digital Knowledge, Department of Classical Philology and Italian Studies, University of Bologna, Bologna, Italy

[2] Research Centre for Open Scholarly Metadata, Department of Classical Philology and Italian Studies, University of Bologna, Bologna, Italy

[3] Digital Humanities Advanced Research Centre (/DH.arc), Department of Classical Philology and Italian Studies, University of Bologna, Bologna, Italy

**Abstract.** This study focuses on analysing the coverage of publications' metadata available in the Current Research Information System (CRIS) infrastructure of the University of Bologna (UNIBO), implemented by the IRIS platform, within an authoritative source of open research information, i.e. OpenCitations. The analysis considers data regarding the publication entities alongside the citation links. We precisely quantify the proportion of UNIBO IRIS publications included in OpenCitations, examine their types, and evaluate the number of citations in OpenCitations that involve IRIS publications. Our methodology filters and transforms data dumps of IRIS and OpenCitations, creating novel datasets used for the analysis. Our findings reveal that only 36% of IRIS is covered in OpenCitations, with journal articles exhibiting the highest coverage. We identified 5,129,406 citation links pointing to UNIBO IRIS publications. From a purely quantitative perspective, comparing our results with broader proprietary services like Scopus and Web of Science reveals a comparable quantitative coverage in the number of IRIS bibliographic resources included in all the systems analysed (OpenCitations, Scopus and Web of Science) as well as in the number of citations received by them.

# Introduction

The importance of having available *research information*, i.e. metadata that enables one to understand how research is conducted and communicated, is central to several activities that involve research-performing institutions and funding organisations, which include strategic prioritisation, policy decisions, and research outcomes. Recently, there has been much pressure, usually from the academic community and advocates for Open Science practices, to convince the producer of such information to release it as open material to maximise its reuse and, thus, foster transparency for the activities mentioned above. Indeed, in the past few years, we have seen great attention to this respect in official international reports, such as the Recommendation on Open Science by UNESCO (2021), and several initiatives born from scholars such as the Initiative for Open Citations (I4OC, https://i4oc.org), the Initiative for Open Abstracts (I4OA, https://i4oa.org), and CoARA (Coalition for Advancing Research Assessment, 2022).

These attempts have either framed the problem of the availability of the research information into a bigger picture, as by UNESCO (for Open Science) and CoARA (for research assessment), or focused the discussion on specific types of research information, as in I4OC (for open citations) and in I4OA (for open abstracts). However, in 2024, the Barcelona Declaration on Open Research Information (2024a) (DORI, https://barcelona-declaration.org) put research information as the primary focus of its activities. Created with the effort of several parties coming from and/or working with academia and already advocated in the production and publication of open research information, the Barcelona Declaration aimed to gather supporters for a critical agenda organised into four main commitments:

- *openness* – that should be the default for the research information used and produced by research-performing organisations and funders;
- *collaboration* – pushing for working with services and systems that support and enable (i.e. by producing and publishing) open research information;
- *sustainability* – by supporting, e.g. financially and taking part in their governance, open infrastructures dedicated to the production and publishing of open research information;
- *transition* – taking part in collective actions to accelerate the transition to openness of research information.

An initial agenda of the priorities co-created by the Declaration's signatories and supporters has been published as one of the outcomes of the Paris Conference on Open Research Information, held in September 2024. The agenda is derived from the conference report (Barcelona Declaration on Open Research Information, 2024b), where the primary and most voted action item was dedicated to replacing closed systems (e.g. Web of Science and Scopus) with open alternatives. Of course, other actions have been highlighted as very important as well, as they are crucial prerequisites for implementing such a replacement, in particular the evaluation of such existing open data sources on their quality, coverage (in terms of both kinds of research outcomes represented and additional contextual information such as fundings and grants), and openness/transparency aspects.

However, it is essential to highlight that, even before the Barcelona Declaration, the use of providers for open research information has already been a well-known practice in recent years by several communities, including the scientometrics domain, notably to support the fundamental shift toward more transparent and reproducible analysis. OpenAIRE (Manghi et al., 2012) stands out as a prominent example that has been used for studies to determine potential alternatives to commercial sources for research assessment exercises in Italy (Bologna et al., 2022), for assessing the quality and utility of OpenAIRE in monitoring EU-funded research (Mugabushaka et al., 2021), and for enabling scientometric analysis using OpenAIRE data (Mannocci & Baglioni, 2024). Also OpenCitations (Peroni & Shotton, 2020) is an open scholarly infrastructure dedicated to the publication of bibliographic metadata and citation data that have been used in several studies within the scientometrics community to analyse particular aspects of scholarly phenomena such as retractions in the Humanities (Heibi & Peroni, 2022), citations to books (Zhu et al., 2019), and impact of research grants in Brazil (Perlin et al., 2024). As a last example, OpenAlex (Priem et al., 2022) is another relevant provider that has been used in several recent studies to examine its coverage compared to commercial alternatives (Culbert et al., 2025; Maddi et al., 2025; Céspedes et al., 2025) or to investigate other phenomena, such as scholarly retractions (Ortega & Delgado-Quirós, 2024) and bibliometric analysis (Alperin, 2024).

Even if, in recent years, the data from open scholarly infrastructures has been already used in research studies, including those highlighted above, there has not been a broad evidence of adoption of such open research information in the context of institutional activities and commitments (e.g. in research assessment exercises), which is one of the primary mandates of the Barcelona Declaration, in particular in some countries such as Italy. Indeed, this is an area that still needs appropriate experimentation involving directly the Declaration's signatories to demonstrate, for instance, that the open research information already available in several providers (such as OpenAIRE, OpenCitations, and OpenAlex) to describe research products (e.g. publications) of an institution has a comparable coverage with that coming from proprietary services. The more coverage studies we have to confirm this hypothesis, the easier is to convince institutions to work closely with (and move investments to sustain) open research information providers to improve data quality and coverage and, eventually, to replace closed systems with open metadata also in institutional settings – goals that are aligned with the roadmap of the Declaration (as of 3 August 2025).

Within this scenario, the University of Bologna, one of the initial signatories of the Declaration, is working to devise possible paths to comply with all the Declaration's commitments. The work presented in this paper introduces part of the effort at the University of Bologna to analyse fundamental dimensions related to such commitments, focusing on understanding the requirements necessary to meet, in principle, the commitment *transition* of the Declaration. In particular, the research questions (RQs) we address in this work are the following:

1. What is the current coverage (in terms of number and publication type) of the publications authored by a scholar affiliated with the University of Bologna (UNIBO publications from now on) in an existing and well-recognised source of open research information, i.e. OpenCitations?
2. According to OpenCitations, how many citation links involve UNIBO publications (either as citing entity, cited entity, or both)?

To answer these questions and make the whole analysis transparent and reproducible, many requirements, complying with the Declaration's commitments, had to be met. First, we needed access to all bibliographic metadata of UNIBO publications available under open licenses and published using open and machine-readable formats (commitment *openness*). That has been addressed thanks to the collaboration of two units of the University of Bologna, dedicated to IT Systems and Services (CeSIA) and Planning and Communication (APPC), which enabled us to produce a CSV dataset with all the UNIBO publications (as of 30 May 2025) to use for the analysis.

Second, we needed to work with one of the authoritative sources of open research information containing bibliographic metadata and citation data to measure the data coverage highlighted in RQ1 and RQ2 (commitment *collaboration*). We chose to interact with OpenCitations (https://opencitations.net) (Peroni & Shotton, 2020), which is an independent not-for-profit infrastructure organisation dedicated to the publication of open bibliographic and citation data that is managed, for administrative purposes, by the Research Centre for Open Scholarly Metadata (https://openscholarlymetadata.org) of the University of Bologna (*sustainability* commitment).

Given these premises, the results obtained from our analysis sketch out an initial picture of the current status of alignment with open research information providers and set up possible paths for further studies and experimentation. In addition, as a direct consequence of the study, it has initiated a practice of publishing yearly dumps of bibliographic information of all UNIBO publications into the University's institutional repository for research data (AMSActa, https://amsacta.unibo.it).

The rest of the paper is organised as follows: In Section "Materials and methods", we introduce all the data, protocols, and methodology developed for running the analysis. In Section "Results", we present the outcomes of our analysis, which are then discussed mainly in Section "Discussion". Finally, in Section "Conclusions", we conclude the paper and sketch out some future work.

# Materials and methods

This section introduces all the data and protocols adopted for the analysis. All materials produced (Zilli et al., 2025a-g) are available online to enable the reproducibility of the study. More details about these resources can be found in Section "Data Availability Statement".

## Data reused

The analysis proposed in this paper reuses data included in two different sources: one institutional source, the Institutional Research Information System (IRIS) used by the University of Bologna, which contains metadata about UNIBO publications, and an open science infrastructure providing bibliographic metadata and citation data, i.e. OpenCitations, having a broader scope in coverage worldwide.

### IRIS

The IRIS software system (Bollini et al., 2016) has been developed by CINECA, a not-for-profit Consortium comprising 70 Italian universities, 4 Italian Research Institutions, and the Italian Ministry of Education. Most Italian universities adopt this software to handle their current research information system (CRIS). IRIS enables universities to collect and organise the bibliographic metadata of all the institutions' scientific production, allows scholars' direct involvement in providing information about their products, and uses a generic data model shared by all IRIS installations – one for each institution involved.

IRIS is used by the University of Bologna, which organises yearly campaigns (with a deadline set to the end of February) to update the status of the related database of research products massively, even if scholars can update their publication information when preferred during the year. The information contained in IRIS concerns basic bibliographic metadata about scientific products (title, author list, publication year, publication venue, persistent identifiers, etc.) but also contains metadata bound to specific licenses and agreements with the publishers (e.g. the abstract of the scholarly articles) and personal data (e.g. the name of the people who have worked on the curation of such metadata) that cannot be shared with licenses enabling to maximise their reuse such as Creative Commons Zero (CC0, https://creativecommons.org/publicdomain/zero/1.0/).

To this end, we have worked on a dump of the University of Bologna's IRIS data provided at the beginning of May 2025, i.e. taken immediately after the last massive update at the University, to extract only the relevant metadata we could publish safely using CC0 as a license. The dataset produced and used in this work (Amurri et al., 2025) is hosted in the AMSActa Institutional Research Repository. It comprises bibliographic metadata of all UNIBO publications (research articles, books, databases, etc.) available in IRIS.

To better understand the time coverage of the dataset and the distribution of records over the years, Figure 1 presents the annual count of publications included in IRIS from 1953 to 2025 (incomplete). This visualisation highlights the temporal span of the data and shows how publication numbers vary by year. The starting point, in 1953, reflects the earliest publication year found in the dataset. The increase shown starting from 2004 relates to a policy related to the introduction of a nationwide research assessment exercise for the universities and other research institutions called *Valutazione della Qualità della Ricerca* (VQR, in English: *Research Quality Evaluation*, https://www.anvur.it/en/research/evaluation-research-quality). While it is still run every five years, the first edition of this assessment exercise was run in 2011 and considered all the publications done by Italian research institutions from 2004 to 2010. Indeed, IRIS was introduced at the national level to simplify the transferring of the publication metadata of universities to the national research metadata collection portal used for such research assessment exercises and managed by the *Italian National Agency for the Evaluation of Universities and Research Institutes* (in Italian: Agenzia Nazionale di Valutazione del Sistema Universitario e della Ricerca, or ANVUR – https://www.anvur.it/), which is controlled by the Italian Ministry of University and Research.

The relatively low number of records for 2025 is instead explained by the fact that the data dump only includes publications stored by researchers by the beginning of May 2025. It therefore does not represent the full annual output.

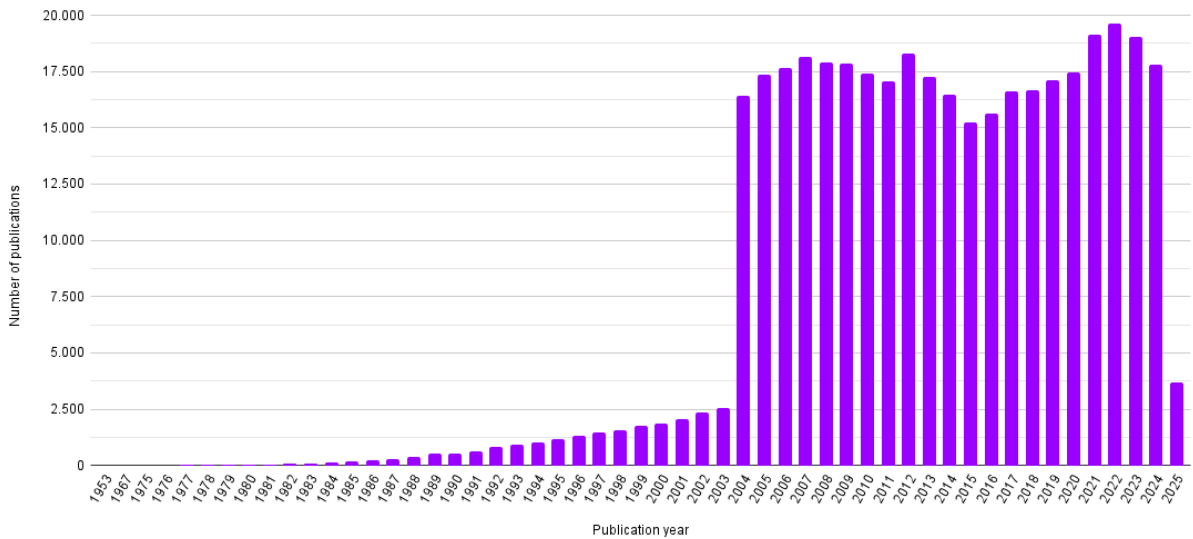

*Figure 1. The number of publications per year of publication in IRIS. The considered range spans from 1953 to 2025 (incomplete).*

This dataset contains seven distinct CSV files, each describing a specific aspect of the publications, with a total of 402,505 bibliographic records. As summarised in Table 1, it includes details about the people involved, such as authors and editors, and publication identifiers like DOIs. Additionally, the dataset captures information on the language of the publications, basic descriptive entities such as titles, publication dates, and types, along with publisher information. In addition, a README file accompanies the dataset, offering additional documentation and guidance.

As described in the documentation of the schema used for each column in the various files (available online in Italian at https://wiki.u-gov.it/confluence/display/public/UGOVHELP/ODS+-+IR-L1), not all the metadata attributes are mandatory. When information is not provided, the value in the CSV files is left blank.

*Table 1. A description of the CSV files included in the IRIS dataset dump.*

| Filename | Description |
| --- | --- |
| ODS_L1_IR_ITEM_CON_PERSON.csv | Information (internal ID, ORCID, name, etc.) about each individual (authors, editors, etc.) involved in the publications of the dataset |
| ODS_L1_IR_ITEM_DESCRIPTION.csv | List of authors and author count for each publication |
| ODS_L1_IR_ITEM_IDENTIFIER.csv | Identifiers of publications, such as DOIs, PMIDs, ISBNs and others |

| Filename | Description |
| --- | --- |
| ODS_L1_IR_ITEM_LANGUAGE.csv | Language of the publication (if applicable) |
| ODS_L1_IR_ITEM_MASTER_ALL.csv | Basic metadata (title and publication date) |
| ODS_L1_IR_ITEM_PUBLISHER.csv | The names and locations of the publishers of the BRs |
| ODS_L1_IR_ITEM_RELATION.csv | Additional metadata regarding the publication context (venue, editors, etc.) |

## OpenCitations

OpenCitations (Peroni & Shotton, 2020) is a community-guided open infrastructure that provides access to global scholarly bibliographic and citation data. The infrastructure offers its data for bulk download and enables programmatic access via various interfaces, including REST APIs and a Web GUI. OpenCitations uses Semantic Web technologies to model citations and bibliographic metadata (Daquino et al., 2020), providing comprehensive, freely accessible data (under a CC0 license) while ensuring semantic interoperability.

OpenCitations manages and maintains two main collections, both relevant to the purposes of this work. The first collection is the OpenCitations Index (OC Index from now on) (Heibi et al., 2024), a unified repository of open citations aggregated from various sources – Crossref (Hendricks et al., 2020), DataCite (Brase, 2009), National Institute of Health - Open Citation Collection (Hutchins et al., 2019), OpenAIRE (Manghi et al., 2012), and the last source ingested, i.e. the metadata made available by the Japan Link Centre (Moretti et al., 2024). The second collection, OpenCitations Meta (OC Meta from now on) (Massari et al., 2024), comprises the bibliographic metadata of all citing and cited bibliographic resources included in the OpenCitations Index. Provenance data and change tracking information are also generated for both collections using a provenance model based on the PROV Ontology (Lebo et al., 2013). This approach ensures transparency and traceability by capturing detailed information about data creation/modification, actors involved, and primary sources.

The OC Meta dump used in this work was published in June 2025 (OpenCitations, 2025b). Each line in the CSV dump represents a bibliographic entity and its corresponding metadata. These metadata fields provide information such as the document's unique ID(s) (DOI, PMID, ISSN, OpenAlex ID, etc.), title, authors, publication date, and the venue to which the document has been published, if any. In addition, details about the journal issue and volume are tracked (if applicable), page ranges are recorded, and the type of resource, the publisher, and any editors involved are noted. This structured metadata allows for a comprehensive basic description of each bibliographic entry.

All bibliographic entities in OC Meta are identified using an internal identifier called OMID (the OpenCitations Meta Identifier). The OMID structure is as follows:

[entity_type_abbreviation]/[supplier_prefix][sequential_number]

For example, *br/0601* is a valid OMID, where *br* stands for *bibliographic resource*, *060* is one of the supplier prefixes indicating the dataset (i.e. OC Meta), and *1* is the sequential number.

The dataset contains 124.5M bibliographic entities. Table 2 shows an example of how an entity is represented in the OC Meta CSV dump. As shown in Figure 2, the total number of publications in OpenCitations Meta is reported for each year of publication. It is worth noting that, in Figure 2, we have chosen to report bibliographic entities published from 1953, aligning with those in Figure 1, despite OC Meta containing some entities with publication dates prior to 1953. In addition, the number of publications listed for 2025 is drastically lower than 2024 since the 2025 records are primarily based on the Crossref dump harvested at the beginning of April, thus including records published by the end of March 2025.

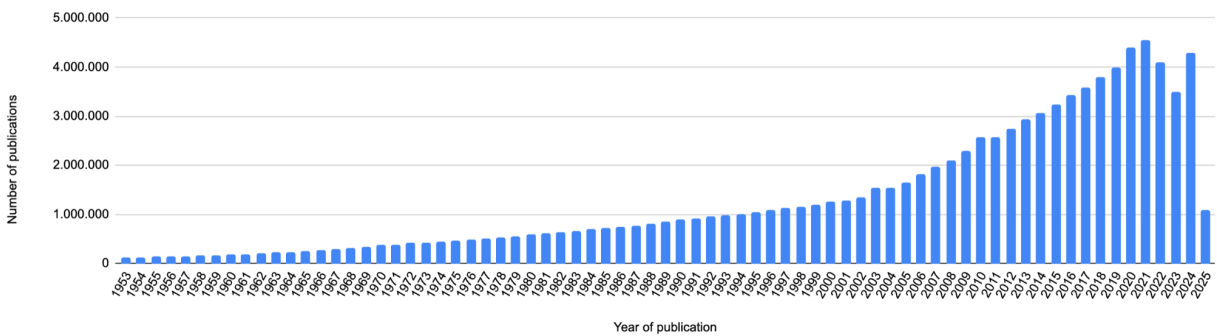

*Figure 2. The number of publications per year of publication in OpenCitations Meta. The considered range shown in the graph spans from 1953 to 2025 (incomplete).*

*Table 2. A sample taken from the OC Meta CSV dump; the first column represents the attributes (columns in the CSV) of the corresponding bibliographic entity.*

| Attribute | Value |
|---|---|
| id | doi:10.1007/978-3-030-00668-6_8 openalex:W2891148407 omid:br/061602192186 |
| title | The SPAR Ontologies |
| author | Peroni, Silvio [orcid:0000-0003-0530-4305 omid:ra/0614010840729]; Shotton, David [omid:ra/061606526499] |
| issue | |
| volume | |

| Attribute | Value |
|---|---|
| venue | The Semantic Web – ISWC 2018 [doi:10.1007/978-3-030-00668-6 isbn:9783030006679 isbn:9783030006686 openalex:W4240995052 omid:br/0611064361] |
| page | 119-136 |
| pub_date | 2018 |
| type | book chapter |
| publisher | Springer Science And Business Media Llc [crossref:297 omid:ra/0610116006] |
| editor | Vrandečić, Denny [orcid:0000-0002-9593-2294 omid:ra/0617010445012]; Bontcheva, Kalina [omid:ra/061408185630]; Suárez-Figueroa, Mari Carmen [omid:ra/061408185631]; Presutti, Valentina [omid:ra/061408185632]; Celino, Irene [orcid:0000-0001-9962-7193 omid:ra/0616010539120]; Sabou, Marta [orcid:0000-0001-9301-8418 omid:ra/0625037023]; Kaffee, Lucie-Aimée [orcid:0000-0002-1514-8505 omid:ra/06160100340]; Simperl, Elena Paslaru Bontas [orcid:0000-0003-1722-947X omid:ra/061409214] |

The OC Index dump used in this work was published in July 2025 (OpenCitations, 2025a). Each line in the CSV dump represents a citation, treated as a first-class data entity, each with its own specified metadata. Such metadata includes:

1. the identifier of the citation;
2. the citing entity;
3. the cited entity;
4. the citation creation date (corresponding to the publication date of the citing entity)
5. the time interval between the citing and cited publication dates;
6. a yes/no flag indicating whether it is an author self-citation (i.e. when the citing and cited entities share at least one author);
7. a yes/no flag indicating whether it is a journal self-citation (i.e. both citing and cited entities are published in the same journal).

Each citation is identified using an Open Citation Identifier, or OCI (Peroni & Shotton, 2019). The OCI structure of the citations in the OC Index is as follows:

```
oci:[citing_n_omid]-[cited_n_omid]
```

For example, *oci:06101801781-062501777134* is a valid OCI, where *06101801781* is the numeral part of the OMID of the citing bibliographic resource (i.e. *br/06101801781*), and *062501777134* is the numeral part of the OMID of the cited bibliographic resource (i.e. *br/062501777134*).

In Table 3, we show an example of how the citation *oci:06404659278-06201483429* and corresponding attributes are represented in the CSV dump. The dump contains 2,216,426,689 unique citations and weighs 38.8 GB when zipped (242 GB unzipped).

*Table 3. A sample taken from the OC Index CSV dump. The first column represents the attributes (columns in the CSV) of the corresponding citation. The citation timespan is represented using the duration XSD datatype (Biron & Malhotra, 2004) having the shape "PnYnMnD", where "P" indicates the period, "nY" indicates the number of years, "nM" indicates the number of months, and "nD" indicates the number of days.*

| Attribute | Value |
| --- | --- |
| id | oci:06404659278-06201483429 |
| citing | omid:br/06404659278 |
| cited | omid:br/06201483429 |
| creation | 2023-11-29 |
| timespan | P2Y3M12D |
| journal_sc | yes |
| author_sc | no |

## Methodology

To answer our two research questions, we defined a methodology that uses the two datasets described in the previous subsection as input to compare bibliographic data in IRIS against those in OpenCitations. The methodology is summarised in the workflow diagram shown in Figure 3. The workflow comprises five steps, each managed by a dedicated tool (graphically represented by a circle with an engine icon). The diagram also includes the numerical outcomes of each step. We provide a step-by-step explanation of the workflow, detailing the processes involved and the output obtained at each stage.

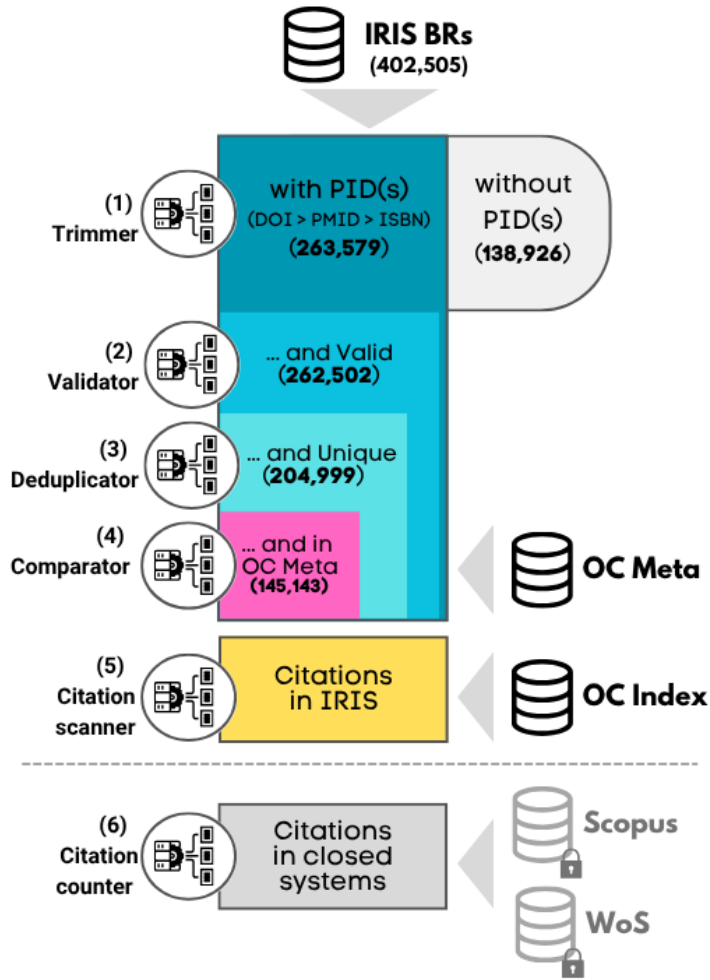

*Figure 3. Overview of the workflow for the adopted methodology. The workflow consists of six steps, beginning with the entire collection of bibliographic resources from IRIS, refining the data, and culminating in a comparison with the OC Meta and Index datasets. A final step also includes a quantitative comparison with research information coming from two proprietary services, i.e. Scopus and Web of Science. Output numbers for the first four steps are also reported.*

## Trimmer

In this initial stage, we collect all bibliographic entities indexed in IRIS and categorise them based on whether they are associated with persistent unique identifiers (PIDs) from the IRIS dataset. Precisely, we extract entities in IRIS, including DOI, ISBN, or PMID identifiers, as these are the only ones present in the IRIS dump and OC Meta.

Of the 402,505 entities in the IRIS dump, we found 263,579 with at least one of these PIDs. From this filtering process, we created the first of our novel datasets, *Iris No ID* (Zilli et al., 2025f), containing the metadata of 138,926 IRIS entities without a DOI, ISBN, or PMID. The remaining

263,579 entries with at least one of the PIDs supported in OC Meta are used to build the list of unique identifiers.

## Validator

For each bibliographic resource (BR from now on) in IRIS, we select one of the three PIDs, prioritising DOIs, PMIDs, and ISBNs. We developed a heuristic that prioritises DOIs and PMIDs because they directly identify content, such as articles and datasets, which are central to our dataset. ISBNs, i.e. identifiers for books that often serve as containers (i.e. the venue) of aggregated knowledge, are considered only as a fallback when content-specific identifiers are unavailable. In cases in which one BR has more than one identifier available, the first is picked. Malformed identifiers are sanitised (e.g. removing the leading zeros from PMIDs and removing hyphens and spaces from ISBNs), and syntactically invalid ones are discarded by extracting only the identifiers with valid patterns using regular expressions, as exemplified in Table 4. All identifiers are normalised following the OC naming convention used in the OC Meta CSV files – *prefix:identifier*, where *prefix* indicates the identifier type (e.g. `doi`, `pmid`, `isbn`) and *identifier* is the literal string of the identifier in lowercase. Table 5 summarises the number of identifiers before and after the filtering, validation, and normalisation process.

*Table 4. Examples from the filtering, validation, and normalisation process. All issues recognised during the validation, which lead to the discarding of an identifier, are underlined.*

| PID Type | BR IDs in IRIS | Discarded/normalised |
|---|---|---|
| DOI | 10.3303/CET1543057 | doi:10.3303/cet1543057 |
| | 10.193/infdis/jiu617 | discarded |
| | 9788838697340 | discarded |
| PMID | PMID: 9276009 | pmid:9276009 |
| | PMC 4874964 | discarded |
| | PMC2206475 | discarded |
| ISBN | 888809556X; 978-8888095561 | isbn:888809556x |
| | 88.6080.002.1 | discarded |
| | (OBRA COMPLETA):; (VOL. I) | discarded |

*Table 5. Number of entities with DOI, ISBN, and PMID before and after the validation process.*

| PID Type | IRIS count | Invalid PID count | Valid PID count |
|---|---|---|---|
| DOI | 184,454 | 238 | 184,216 |
| PMID | 59,984 | 6 | 59,978 |

| PID Type | IRIS count | Invalid PID count | Valid PID count |
|---|---:|---:|---:|
| ISBN | 93,775 | 974 | 92,801 |
| *Total* | *338,213* | *1218* | *336,995* |

## Deduplicator

The output of the previous step undergoes a process of deduplication to remove the 86,306 cases in which we found the same PID associated with multiple IRIS entries. This situation may be the result of different scenarios, including either the production of duplicated records, e.g., when two distinct UNIBO authors add to IRIS the same entity twice, or the specification of the same identifier for a bibliographic resource and its venue (e.g., the same ISBN specified to a book and to all the chapters it contains).

To address these instances, we act on DOIs, ISBNs and PMIDs separately, establishing a priority system that ranks the duplicated BRs based on their type and allows us to pick the preferred one. This system, implemented according to the priorities of the types of bibliographic resources described in Tables 6, 7 and 8 (the lower number, the bigger priority), has been devised following the manual investigation of sample duplicate records, which led to the discovery that only selected types ensure that the final dataset includes only the most relevant and accurate entries filtered between PIDs..

The approach works as follows. First, we gather all the BRs having the same PID specified and use metadata from the original IRIS dataset to count the number of missing values in each record. Then, within groups sharing the same type and PID, we sort the records by the number of non-null fields in descending order and keep only the most complete record per subgroup. Next, for the remaining records of the group sharing the same PID but different types, we sort them according to the priority number determined for the BR types, and select the first out of the ordered records. This method allows us to deduplicate BRs that share the same type, as well as BRs with different types. For instance, if we have three entities with the same DOI, and two have been defined as journal articles and the other as a proceeding article, we deduplicate them as a single entity, choosing the journal article as its final type, having better priority than the proceeding article. The cases addressed with these priority tables mainly involved multiple entries referring to the insertion of the same PID for both content (e.g. a book chapter) and container/venue (e.g. a book). By the end of this process, we are left with a list of 204,999 unique PIDs, reduced from the original 262,502 PIDs, as described in Table 9.

*Table 6. DOI priority table. The first column indicates the related BR type we used for alignment purposes in OC Meta. In contrast, the second column lists the IRIS type specified in the IRIS dataset (with its Italian label). A lower number (third column) has a higher priority for that type.*

| OC Meta type | IRIS type | Priority |
|---|---|---|
| Journal article | 1.01 Journal Article | 0 |
| Book | 3.02 Edited volume | 1 |
| Book chapter | 2.01 Chapter / Essay in book | 2 |
| Proceedings article | 4.01 Contribution in conference proceedings | 3 |

*Table 7. PMID priority table. The first column indicates the related BR type we used for alignment purposes in OC Meta. In contrast, the second column lists the IRIS type specified in the IRIS dataset (with its Italian label). A lower number (third column) has a higher priority for that type.*

| OC Meta type | IRIS type | Priority |
|---|---|---|
| Journal article | 1.01 Journal Article | 0 |

*Table 8. ISBN priority table. The first column indicates the related BR type we used for alignment purposes in OC Meta. In contrast, the second column lists the IRIS type specified in the IRIS dataset (with its Italian label). A lower number (third column) has a higher priority for that type.*

| OC Meta type | IRIS type | Priority |
|---|---|---|
| Book | 3.02 Edited volume | 0 |
| Book | 3.01 Monograph | 1 |
| Journal article | 1.01 Journal Article | 2 |

*Table 9. Number of BRs uniquely identified by a DOI, PMID (and not by a DOI), and ISBN (and not by a DOI nor PMID) after deduplication.*

| PID schema | Final Unique PIDs count |
|---|---|
| DOI | 161,455 |
| PMID | 2,119 |
| ISBN | 41,425 |

| PID schema | Final Unique PIDs count |
|---|---:|
| *Total* | *204,999* |

## Comparator

At this stage, we extract the data from the OC Meta dataset. We look for the BRs in the current collection of IRIS obtained in the previous passage in the OC Meta dataset. During this process, we identified 1,121 cases where unique IRIS BRs appear two or more times in OC Meta, where the same IRIS BR appears multiple times with different OMIDs. We remove these duplicates, keeping the occurrence with the most complete set of information available.

In order to ensure temporal comparability between the OC Meta dataset and the IRIS dataset, we set a cutoff period for the extraction of the BRs. Specifically, we retain only BRs whose publication year is less than or equal to 2025. If the publication date in OC Meta is missing, the corresponding publication year present in IRIS is used as a fallback for the temporal filtering. Since the IRIS dataset was extracted at the beginning of May 2025, we assume that no resources in it should have a publication date after this year. An inspection of the filtered results confirmed that this temporal filtering only excluded 13 records, all of which had either a missing or invalid (e.g. "9999") publication date in both IRIS and OC Meta.

The final output of this process consists of the two primary datasets used in our study, *Iris in Meta* (Zilli et al., 2025e) and *Iris Not in Meta* (Zilli et al., 2025g), which contain all deduplicated IRIS BRs that are included and not included in OC Meta, respectively.

## Citation Scanner

In the final stage, we run a preliminary citation analysis of the IRIS BRs included in *Iris in Meta*. In particular, we extract all OMIDs from the *Iris in Meta* dataset and filter the OC Index to identify all citations where an OMID from the list appears as either the citing or cited entity. We apply the same publication year cutoff used in the creation of the previous datasets, retaining only citations involving citing BRs published in or before 2025. The result of this process is a new dataset, *Iris in Index* (Zilli et al., 2025d).

## Citation Counter

In addition, to compare the citation extracted from OpenCitations with other proprietary citation indexes, we asked to the Planning and Communication unit (APPC) of the University of Bologna, which is responsible of handling IRIS and tracking metrics for IRIS publications, to provide us the citation counts by Scopus (Baas et al., 2020) and Web of Science (Birkle et al., 2020), on the 1st of April 2025, of all the IRIS entities included in our dataset. These data enable us to quantitatively compare the data extracted from OpenCitations with well-known proprietary sources, and to measure the average number of citations per publication to assess their global comparability. In particular, the Scopus and Web of Science snapshots have been created through querying the related API services made available by the two systems, which included all the indexes in the Core Collection for what concerns Web of Science.

# Results

In this section, we present and highlight key results derived from the analysis of the four datasets produced by running our methodology: *Iris No ID*, *Iris in Meta*, *Iris Not in Meta*, and *Iris in Index*. These results provide critical support for our discussion of addressing the research questions RQ1-RQ2 introduced in Section "Introduction".

## Bibliographic Records types

The number of records in *Iris in Meta* amounts to 145,143, representing 70.8% of the list of deduplicated PIDs extracted from IRIS and 36% of the unfiltered IRIS dump (Amurri et al., 2025). The complete breakdown of IRIS publication types as included in the deduplicated IRIS dataset and *Iris in Meta* is provided in Appendix [1](#), showing that journal articles have the highest coverage rate (91.6%), followed by brief publications (78.5%), which shows a minimal number of items in IRIS, and conference proceedings (53.1%).

Comparing the types of BRs as described in IRIS and OC Meta, based on the data in *Iris in Meta*, also reveals some interesting insights on the extent to which the types align between the two systems. We analysed the types of IRIS BRs in *Iris in Meta* and the corresponding OC Meta types resulting from the type alignment we performed (introduced in Appendix [2](#)). Then, we checked the number of BRs that respected such an alignment according to the type we retrieved from OC Meta. This analysis shows that most BRs (134,088) have a coherent type between the two datasets, while a smaller portion (11,055 BRs) shows a different BR type in IRIS and OC Meta. As shown in Figure [4](#), there is a more heterogeneous and diversified distribution for types when the number of BRs of that IRIS type is significant. In contrast, perfect matches are observed only for specific types, such as series, computer programs, and book series, often represented by a single item.

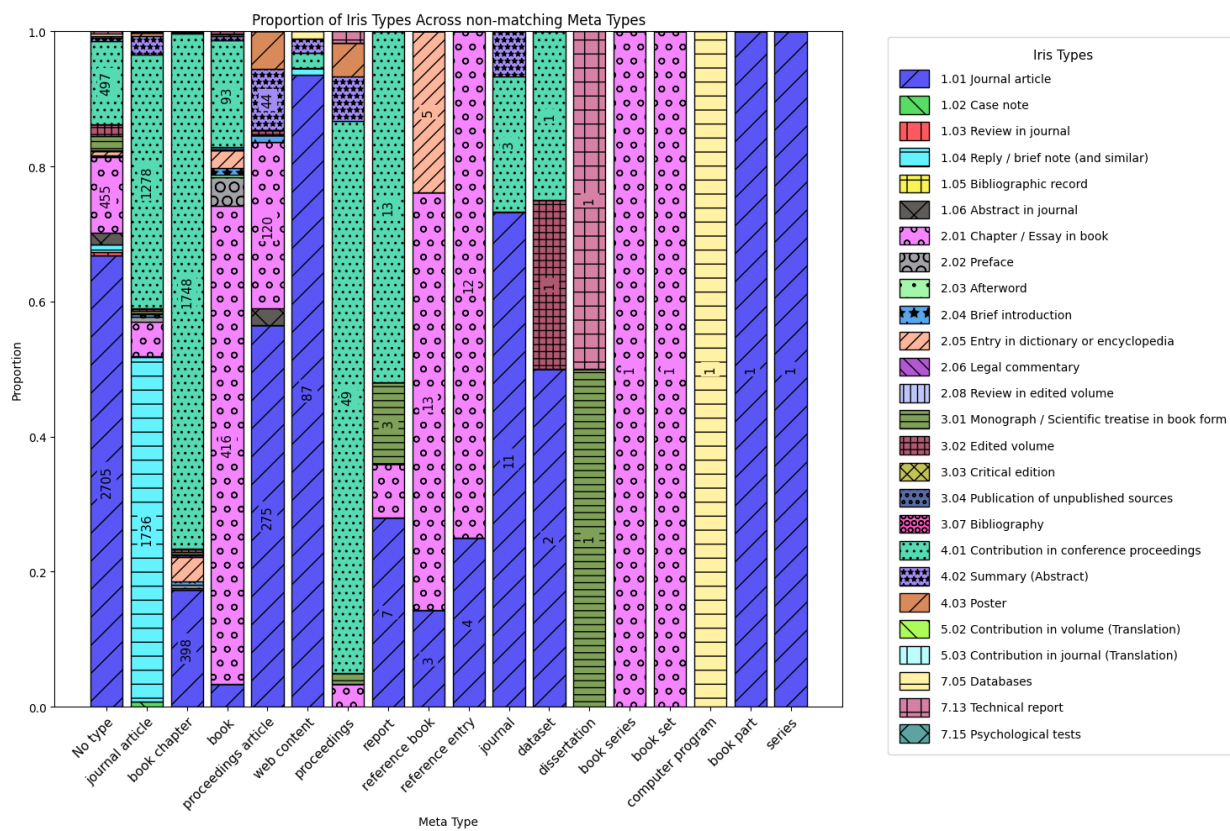

*Figure 4. Overview of the mismatching types between IRIS and OC Meta.*

## Bibliographic Records without Permanent IDs

Analysing *Iris No ID*, we found that 34.5% of the BRs in the original IRIS data dump (138,926 records) do not have any of the PID schemes that OC Meta collects in their databases. The distribution of entries in this dataset, shown in Table 10, highlights a predominance of journal-based products, with relatively fewer contributions for books and reviews.

*Table 10. Overview of the top 5 types of BR in the Iris No ID dataset.*

| IRIS type | count |
|---|---:|
| 1.01 Journal Article | 55,368 |
| 4.02 Summary (Abstract) | 18,781 |
| 4.01 Contribution in conference proceedings | 16,849 |
| 2.01 Chapter / Essay in book | 10,271 |
| 1.03 Review in journal | 6,158 |

To better understand the reasons behind this low coverage, we conducted a dedicated study on the *Iris No ID* dataset (Andreose & Zilli, 2025). The study involved the cleaning and deduplication of the IRIS records lacking DOIs, ISBNs, or PMIDs, followed by a metadata-driven analysis of publication type, country, and year. Temporal patterns revealed that missing PIDs were particularly frequent in entries from 2004-2008, likely due to a bulk migration from an older system used by the University of Bologna to store bibliographic information, with a steady improvement observed in subsequent years, when the use of IRIS for storing bibliographic metadata became a common (and mandatory) practice at the University.

To explore possible enrichment strategies, we implemented an automated retrieval pipeline using Crossref's public API (https://api.crossref.org). By querying each BR via a combination of title and authors, we were able to successfully reconcile 18,664 records (18% of the dataset) with Crossref DOIs. These were subsequently cross-checked with OC Meta, where 10,387 matches were found. Notably, 5,754 of the retrieved DOIs were associated with multiple IRIS entries, suggesting the presence of metadata inconsistencies or unresolved duplicates within the original dataset. Addressing these issues, through more robust disambiguation and record consolidation, will be essential in future efforts to improve the reliability and interoperability of IRIS data.

## Bibliographic Records not included in OpenCitations Meta

From *Iris Not in Meta*, we discovered that 29.1% of the deduplicated PIDs extracted from IRIS and searched within OC Meta did not find a match. The vast majority of this portion of BRs comprises entities for which only the ISBN identifier was found while extracting the list of PIDs from IRIS. We identified 41,141 ISBNs, 17,585 DOIs, and 1,130 PMIDs that are not present in OpenCitations Meta. Table 11 provides a detailed breakdown of the five most frequently occurring types of BRs missing from OC Meta.

Table 11. Overview of the top 5 types of the IRIS BRs not present in OC Meta.

| IRIS type | count |
| --- | --- |
| 2.01 Chapter / Essay in book | 19,332 |
| 1.01 Journal article | 11,474 |
| 4.01 Contribution in conference proceedings | 8,935 |
| 3.01 Monograph / Scientific treatise in book form | 7,064 |
| 3.02 Edited volume | 5,628 |

## Duplicated Bibliographic Records

While extracting the list of identifiers from IRIS, we identified cases where IRIS entries were associated with more than one PID. Specifically, 86,306 duplicates were found, involving 28,803 PIDs linked to more than one IRIS entry. This issue is particularly prevalent with ISBNs, as most duplications stem from incorrect aggregation of content and container identifiers.

Indeed, it is common in IRIS to find entries for distinct items sharing the same PID, where the PID does not explicitly refer to any of the individual items but rather to the larger container (i.e. the venue) that holds them. A clear example of this issue is the case of 285 IRIS entries, which represent a series of individual entries from the *Dizionario Biografico degli Italiani*, all linked to the same ISBN of the volume of the dictionary in which they are contained. This pattern is frequently observed in similar cases involving dictionary or encyclopedia entries, book chapters, proceedings articles, and journal articles. Table 12 summarises the number of duplicate BRs in IRIS grouped by PID schema.

*Table 12. The number of duplicate BRs in IRIS by PID schemas.*

| duplicate PID schema | BR count |
|---|---:|
| ISBN | 47,239 |
| DOI | 38,836 |
| PMID | 231 |
| *total* | *86,306* |

## Bibliographic Records in OpenCitations Index

As derived from *Iris in Index*, the total count of the OC Index citations involving deduplicated IRIS BRs amounts to 9,951,613. Table 13 breaks this number down by counting the number of times IRIS BRs take on the role of citing entities, cited entities, or both.

*Table 13. Count of the citations in Iris in Idex involving IRIS BRs.*

| Role of IRIS BR | Citation count |
|---|---:|
| Citing | 5,281,530 |
| Cited | 5,129,406 |
| Citing and Cited | 459,323 |

We also extract the citation counts received by IRIS BRs retrieved in Scopus and Web of Science on the 1st of April 2025 to have them comparable with the latest OpenCitations Index dump used in this analysis, which is based on the Crossref dump in April 2025. Thus, it covers bibliographic metadata and citations until the end of March 2025. In particular, we obtain citation information from, respectively, 144,940 distinct IRIS BRs that are also included in SCOPUS and 129,823 distinct IRIS BRs that are also included in Web of Science. In particular, as shown in Table 14, the total number of citations in Scopus and Web of Science amounts to 5,225,193 and 4,505,715, respectively. In addition, the average number of citations per publication, computed by dividing the number of citations by the number of IRIS BRs found in each source, is 36.05 (Scopus), 34.71 (Web of Science) and 35.34 (OpenCitations).

Table 14. Comparison of the citation counts between Scopus, Web of Science and OpenCitations.

| Source | IRIS BRs in source | Citations to IRIS BRs | Citations / BRs ratio |
| --- | --- | --- | --- |
| Scopus | 144,940 | 5,225,193 | 36.05 citations per BR |
| Web of Science | 129,823 | 4,505,715 | 34.71 citations per BR |
| OpenCitations | 145,143 | 5,129,406 | 35.34 citations per BR |

# Discussion

According to the IRIS dump analysed, of the overall 402,505 BRs in IRIS, 36% (145,143) are included in OC Meta (RQ1) – which is aligned with the value observed in the two proprietary databases included in the analysis, i.e. Scopus and Web of Science. Two possible factors can explain this partial coverage. On the one hand, it is important to stress that OC Meta only includes bibliographic resources that take part (either as citing entity or cited entity) in citations included in the OC Index. While the latter collection includes more than two billion citation links, OpenCitations does not have all the possible citation links existing in the literature since that information is lacking from the primary sources used for creating the OC Index. For instance, if a BR available in IRIS has no (incoming and outgoing) citation links in the OC Index, that is not included in OC Meta by construction, resulting in a missing value for the present study. However, this issue can be addressed in the future, for instance, by complementing OpenCitations data with those coming from other open sources, such as OpenAIRE (Manghi et al., 2012) and OpenAlex (Priem et al., 2022) for traditional publications and other archives and repositories for different kinds of research outcomes – e.g. Software Heritage (Di Cosmo & Zacchiroli, 2017).

On the other hand, IRIS includes many types of research outputs (as summarised in Appendix 1) that go beyond those usually available in existing (open and closed) bibliographic databases. A few examples are book chapter or essay (IRIS type: 2.01 Chapter / Essay in book), monograph or scientific book (3.01 Monograph / Scientific treatise in book form), curatorship (3.02 Edited volume), legal comment (2.06 Legal commentary), abstract in journal (1.06 Abstract in journal), and databases (7.05 Databases). In this case, a possible path to fulfil this gap is to work systematically with open infrastructures to create, at least, alignments to enable different systems to technically and semantically interoperate and to enable filling the gap in one system (e.g. UNIBO IRIS) with information in another system (e.g. OpenCitations) and vice-versa. Such layers of system interoperability are one of the core pillars studied and investigated by several task forces – and introduced related reports (Corcho et al., 2021; Kakaletris et al., 2023; Nyberg Åkerström et al., 2024) – working on the European Open Science Cloud (EOSC) (Burgelman, 2021), set up by the EOSC Association (https://eosc.eu) in recent years. Recent efforts in this direction have been devised and proposed in the context of the RDA Scientific Knowledge Graphs – Interoperability Framework (SKG-IF) Working Group (https://www.rd-alliance.org/groups/scientific-knowledge-graphs-interoperability-framework-skg-if-wg/), which proposed a set of specifications (https://skg-if.github.io) to simplify the exchange of metadata about research products and their related contextual information.

We have also analysed the number of incoming and outgoing citations that involve IRIS BRs included in OpenCitations Meta (RQ2). The number of citations IRIS BRs receive is particularly important for local and national activities, particularly those related to research assessment exercises. Currently, the platforms adopted at a national level for extracting such information, as required by ANVUR, are the two proprietary services used in this study, i.e. Scopus and Web of Science. The result shown in Table 14 suggests a similar coverage of OpenCitations with the other proprietary database, being close to that observed in Scopus and greater than that in Web of Science. We have also measured the average amount of citations received by IRIS BRs included in Scopus, Web of Science, and OpenCitations – still summarised in Table 14. The table shows that the average number of citations per BR is very similar across the three sources, thus confirming an apparent quantitative similarity across the three systems, at least in the local context considered in this study.

Another interesting point of analysis in this context is the overlap of the citing entities that cite IRIS BRs in the three sources. It would be important to see, for instance, if the coverage of the citing entities involved in each citation pointing to IRIS BRs is similar across the three sources or, instead, is partially overlapping and complementary to each other. However, for running such an analysis, we would need the open availability of the complete citation data from all three sources, thus comprising information about the basic bibliographic metadata of all the citing entities and the actual link between the citing entities and the IRIS BRs. However, this information is only openly available in OpenCitations data since the citations in Scopus and Web of Science are grouped, and only the citation count is available for this study. This situation again stresses the importance of having available open research information, particularly when running comparative studies.

Comparing OC Meta and UNIBO IRIS, we have noticed that 7.6% of the IRIS BRs included in OC Meta (11,055) have different publication types between the two sources. These mismatches could impact the accuracy of our analysis, especially when comparing publication types in OC Meta and IRIS. Further study for such mapping is necessary in the future and should consider the OC Meta documentation

([https://github.com/opencitations/metadata/blob/master/documentation/csv_documentation-v1_1_0.pdf](https://github.com/opencitations/metadata/blob/master/documentation/csv_documentation-v1_1_0.pdf)) for the complete uptake of bibliographic resource types.

A final point worth addressing is how other studies employing proprietary bibliometric databases have tackled challenges similar to those discussed in this work. Generally, when working with closed and proprietary systems, both quantitative and qualitative comparisons become challenging. This difficulty arises primarily from differences in coverage and the availability of metadata provided by the closed platforms under analysis. Achieving accurate matching and meaningful analysis requires full access to the underlying metadata to ensure that the entities being compared are truly equivalent. Moreover, the providers of closed bibliographic data often follow different strategies and policies – such as localised coverage focused on specific countries or institutions – which directly influence the datasets they make available.

A reasonable approach to analysis could involve considering the methodologies and objectives of studies closely related to the one presented in this paper. In particular, several institutional research information systems have been examined through the integration of data from platforms such as Scopus and Web of Science. For instance, van Leeuwen et al. (2016) combined outputs registered in the Dutch CRIS with data from Web of Science to evaluate the impact of university research. Likewise, Ma and Cleere (2019) investigated the coverage of Scopus, Web of Science, and the Output-Based Research Support Scheme (OBRSS) used at University College Dublin (UCD). In the case of UCD, coverage compared to Scopus and Web of Science (WoS) was highly comparable – and in some cases slightly better – particularly for materials in the Social Sciences and Humanities (SSH). A similar disparity in SSH coverage in WoS was also noted by van Leeuwen et al. (2016).

Considering these methodological aspects, we argue that a key contribution of our study lies in its use of open bibliographic data source – i.e. OpenCitations – as the primary basis for analysis, which has demonstrated that, from a pure quantitative point of view analysed in a specific local context (i.e. the CRIS system of the University of Bologna), the numbers obtained do not differ from those obtained from proprietary and closed sources. Thus, our approach and, in general, the use of providers of open research information facilitate meaningful comparison and offer deeper insight into how institutional and CRIS system data are represented and perform across both open and closed environments.

# Conclusions

In this work, we have presented the results of an analysis comparing the publications' metadata contained in the institutional bibliographic database of the University of Bologna, i.e. UNIBO IRIS, with an Open Science infrastructure containing the same kind of open research information, i.e. OpenCitations. The study's main aim has been to check, on the one hand, the current coverage of the IRIS' publications in OpenCitations and, on the other hand, to see the availability of citations for all these matched publications. The results have shown how, potentially and in perspective, open research information systems can be adopted and replace the currently used closed information systems, at least in the context of the University of Bologna.

Further studies, locally (within the University of Bologna) and globally (involving other universities among the signatories of the Barcelona Declaration, for instance), should be performed to confirm this initial speculation. Indeed, the final research question that, in the future and with a coordinated effort across universities, institutions and infrastructures, we would like to answer should be:

> Is the open research information currently available enough to implement the transition from closed to open systems as aimed by the Barcelona Declaration?

To address this issue, we must gather evidence from different institutional and applicative contexts and use as many potential sources of open research information as possible. Indeed, it is unlikely that we will have a unique open research information system in the future with all the necessary metadata to handle the diverse activities and needs of institutions worldwide. Instead, a federation of providers of open research information, coordinated between them and technically/semantically interoperable, may better serve the needs of the scholarly community.

Considering the work presented in this paper, we are planning further activities for the following months. One material produced and used as a consequence of the present work is the publication of the UNIBO IRIS dataset in CC0 to maximise its reuse in several contexts beyond this analysis and to comply with the Barcelona Declaration's *openness* commitment. We aim to keep the dataset updated by releasing future versions of it every year, initially, and then every six months. In addition, the scripts developed for filtering the data from the original IRIS dump to create the current dataset – which avoids the presence of personal information, as explained in Section "Data reused" – will also be tested with other IRIS installations external to the University of Bologna. Indeed, we have already started to collaborate with other Italian institutions to run similar studies soon, since, as mentioned in Section "Materials and methods", IRIS is used by the majority of Italian Universities. This work would allow us to maximise the reuse of the code developed and experimented with in this study, add additional showcases, and facilitate the creation of open metadata dumps about the Italian scholarly publication landscape, all of which supports the Barcelona Declaration goals.

Finally, from the data production perspective, we aim to initiate an active collaboration with OpenCitations, being two of the authors directly involved in this open scholarly infrastructure, to devise strategies and protocols to potentially extend the coverage of IRIS BRs in OC Meta by implementing plugins for ingesting IRIS-compliant data into OpenCitations collections – e.g. by processing all the entities in the dataset *Iris No ID*. The new ingestion workflow recently implemented by OpenCitations (Moretti & Heibi, 2023) enables the creation of components for plugging additional sources of bibliographic metadata and citation data in and can be used to facilitate the processing of these missing data in IRIS. Such components will, in principle, allow any IRIS installation to be interoperable with OpenCitations, thus enabling the increment of coverage of Italian publications within such an Open Science infrastructure.

# Authors' contribution statements

Erica Andreose: Conceptualization, Investigation, Methodology, Validation, Visualization, Writing – original draft, Writing – review & editing

Salvatore Di Marzo: Conceptualization, Investigation, Methodology, Software, Writing – original draft, Writing – review & editing

Ivan Heibi: Resources, Supervision, Validation, Writing – original draft, Writing – review & editing

Silvio Peroni: Conceptualization, Funding acquisition, Resources, Supervision, Validation, Writing – original draft, Writing – review & editing

Leonardo Zilli: Conceptualization, Data curation, Investigation, Methodology, Software, Validation, Writing – original draft, Writing – review & editing

# Acknowledgement

This work has been partially funded by the European Union's Horizon Europe framework programme under Grant Agreement No 101095129 (GraspOS Project).

This study was based on the outcomes of the analysis held during the Open Science 2023/2024 course (https://www.unibo.it/en/teaching/course-unit-catalogue/course-unit/2023/443753) of the Second Cycle International Degree in Digital Humanities and Digital Knowledge (https://corsi.unibo.it/2cycle/DigitalHumanitiesKnowledge) of the University of Bologna, taught by SP. We thank all the people who participated in the workshop organised in the context of the Open Science course who provided insightful feedback and, thus, implicitly contributed to some of the aspects of the present study. Also, we want to thank Elena Giachino and Ennio Misuraca from the Planning and Communication unit (APPC) and Alberto Amurri from the IT Systems and Services unit (CeSIA) of the University of Bologna for having provided the raw UNIBO IRIS we elaborated to produce the IRIS dataset used in this study, and, again, Elena Giachino and Ennio Misuraca for having provided crucial statistics about IRIS BRs in Scopus and Web of Science, including their related citation counts.

# Data Availability Statement

The datasets used and analysed during the current study are freely available on Figshare (https://figshare.com/), Zenodo and AMS Acta. In particular:

**OpenCitations Meta** – Published on June 9, 2025 (10.5281/zenodo.15625651). The dataset comprises 124,526,660 bibliographic entities, 376,295,095 authors, 2,765,927 editors, 1,019,563 publication venues, and 103,928,927 publishers. The compressed data totals 12 GB (49 GB when uncompressed) and is distributed across 38,602 CSV files.

**OpenCitations Index** – Published on July 15, 2025 (10.6084/m9.figshare.24356626.v6). The dataset comprises 2,216,426,689 citations. The compressed data totals 38.8 GB, while the size of the unzipped CSV files is 242 GB.

**UNIBO IRIS bibliographic data dump** – Published on July 9, 2025 (10.6092/unibo/amsacta/8427). This dataset comprises several CSV files that detail the bibliographic data of all publications by UNIBO scholars stored in IRIS, the national platform for bibliographic data storage. The raw dump used to produce this data was harvested at the beginning of May 2025.

**Scopus and Web of Science data** – The citation data used in the study from Scopus and Web of Science have been obtained using their respective APIs by specifying the Scopus ID and the Web of Science ID of each IRIS BR considered, since such identifiers are stored within the IRIS installation of the University of Bologna. The kind of data that we could access via these APIs is regulated by national agreements (signed by the University of Bologna), which allowed us to access only limited information. For instance, while we retrieved citation counts for each of the considered IRIS BRs, we could not identify which entities cited the IRIS BRs. In addition, because of these signed agreements, we are not legally entitled to publish any of the raw data coming from Scopus and Web of Science used in the study, thus preventing its full reproducibility.

The following datasets were produced as part of this study and are openly available:

**Iris in Meta** – Published on August 7, 2025 (10.6084/m9.figshare.25879420.v3). Contains all UNIBO IRIS bibliographic records matched to entries in OpenCitations Meta. The data is stored in Parquet format and totals 7,8 MB.

**Iris in Index** – Published on August 7, 2025 (10.6084/m9.figshare.25879441.v3). Contains all UNIBO IRIS bibliographic records which are found either as citing or cited entities within the OpenCitations Index. The data is stored in Parquet format and totals 177 MB.

**Iris Not in Meta** – Published on August 7, 2025 (10.6084/m9.figshare.25897708.v3). Contains all UNIBO IRIS bibliographic records which are not found within OpenCitations Meta. The data is stored in Parquet format and totals 634 KB.

**Iris No ID** – Published on August 7, 2025 (10.6084/m9.figshare.25897759.v3). Contains UNIBO IRIS bibliographic records lacking any persistent identifiers (DOI, PMID or ISBN). It includes the full range of metadata elements derived from the IRIS dump to facilitate future research on the reconciliation of such records. The data is stored in Parquet format and totals 9 MB.

Additional resources and materials used in this study are openly available:

**Software** – Published on August 7, 2025 (10.5281/zenodo.16759255). All code used for data processing and analysis for the current study is openly available on GitHub (https://github.com/open-sci/2023-2024-atreides-code/) and archived on Zenodo.

**Protocol** – Published on August 7, 2025 (10.17504/protocols.io.g6xmbzfk7). The protocol describing the methodological workflow adopted for this study is available on protocols.io.

**Data Management Plan (DMP)** – Published on August 7, 2025 (10.5281/zenodo.16761351). The updated DMP describing the data policies related to this study is available on Zenodo.

# Conflict of interest

SP is the University of Bologna's representative for the Barcelona Declaration and Director of OpenCitations. IH is the Chief Technology Officer of OpenCitations. OpenCitations is one of the open infrastructures that formally supports the Declaration.

# Appendix 1: BRs type counting in the deduplicated IRIS dataset and Iris In Meta

Table 15. Number of the IRIS types as found in the deduplicated IRIS dataset (IRIS) and IRIS in Meta (Meta), sorted by the percentage of coverage in Meta (%).

| IRIS type | IRIS | Meta | % |
|---|---|---|---|
| 1.01 Journal article | 137,922 | 126,448 | 91.68 |
| 1.04 Reply / brief note (and similar) | 2,270 | 1,782 | 78.50 |
| 4.01 Contribution in conference proceedings | 19,089 | 10,154 | 53.19 |
| 1.06 Abstract in journal | 1,107 | 415 | 37.49 |
| 5.03 Contribution in journal (Translation) | 57 | 17 | 29.82 |
| 7.15 Psychological tests | 7 | 2 | 28.57 |
| 1.02 Case note | 103 | 27 | 26.21 |
| 1.03 Review in journal | 1,237 | 320 | 25.87 |
| 7.05 Databases | 101 | 24 | 23.76 |
| 4.03 Poster | 266 | 54 | 20.30 |
| 2.05 Entry in dictionary or encyclopedia | 783 | 157 | 20.05 |
| 2.01 Chapter / Essay in book | 24,134 | 4,802 | 19.90 |
| 7.13 Technical report | 159 | 31 | 19.50 |
| 4.02 Summary (Abstract) | 1,282 | 172 | 13.42 |
| 2.02 Preface | 642 | 67 | 10.44 |
| 1.05 Bibliographic record | 41 | 3 | 7.32 |
| 2.08 Review in edited volume | 29 | 2 | 6.90 |
| 2.04 Brief introduction | 710 | 43 | 6.06 |
| 3.02 Edited volume | 5,913 | 285 | 4.82 |
| 3.01 Monograph / Scientific treatise in book form | 7,380 | 316 | 4.28 |
| 3.04 Publication of unpublished sources | 47 | 2 | 4.26 |
| 2.06 Legal commentary | 225 | 7 | 3.11 |

| IRIS type | IRIS | Meta | % |
|---|---:|---:|---:|
| 3.07 Bibliography | 46 | 1 | 2.17 |
| 5.02 Contribution in volume (Translation) | 143 | 3 | 2.10 |
| 3.03 Critical edition | 364 | 6 | 1.65 |
| 2.03 Afterword | 155 | 2 | 1.29 |
| 5.01 Book (Translation) | 543 | 1 | 0.18 |
| 2.07 Catalogue entry | 149 | 0 | 0.00 |
| 3.08 Annotated/educational edition | 32 | 0 | 0.00 |
| 3.06 Index | 29 | 0 | 0.00 |
| 5.04 Translation of multimedia or theatrical products | 12 | 0 | 0.00 |
| 7.14 Audiovisual products | 11 | 0 | 0.00 |
| 7.01 Thematic and geographic map | 3 | 0 | 0.00 |
| 3.05 Concordances | 2 | 0 | 0.00 |
| 7.10 Artistic and performance product: Artifact | 2 | 0 | 0.00 |
| 7.11 Artistic and performance product: Art prototype and related projects | 1 | 0 | 0.00 |
| 7.02 Geological map | 1 | 0 | 0.00 |
| 7.03 Prodotto dell'ingegneria civile e dell'architettura | 1 | 0 | 0.00 |
| 8.03 Direction of archaeological excavations | 1 | 0 | 0.00 |

# Appendix 2: IRIS-OC Meta mapping

*Table 16. Mapping of the types of BRs between IRIS OC Meta.*

| IRIS type | OC Meta type |
|---|---|
| 7.05 Databases | dataset |
| 7.15 Psychological tests | |
| 1.04 Reply / brief note (and similar) | other |
| 2.04 Brief introduction | |
| 2.02 Preface | |
| 2.03 Afterword | |
| 4.02 Summary (Abstract) | |
| 4.03 Poster | |
| 1.02 Case note | |
| 2.06 Legal commentary | |
| 3.04 Publication of unpublished sources | |
| 2.05 Entry in dictionary or encyclopedia | reference entry |
| 1.05 Bibliographic record | |
| 4.01 Contribution in conference proceedings | proceedings article |
| 7.13 Technical report | report |
| 2.01 Chapter / Essay in book | book chapter |
| 2.08 Review in edited volume | |
| 5.02 Contribution in volume (Translation) | |
| 3.01 Monograph / Scientific treatise in book form | book |
| 5.01 Book (Translation) | |
| 3.02 Edited volume | |
| 3.03 Critical edition | |
| 1.01 Journal article | journal article |
| 5.03 Contribution in journal (Translation) | |
| 1.06 Abstract in journal | |
| 1.03 Review in journal | |
| 3.07 Bibliography | reference book |
| None | *no type specified* |

# Appendix 3: IRIS Taxonomy: English Mapping

*Table 17. Mapping of IRIS types from Italian (original names) to English*

| IRIS Type (Original – Italian) | IRIS Type (English Translation) |
|---|---|
| 1.01 Articolo in rivista | 1.01 Journal article |
| 1.02 Nota a sentenza | 1.02 Case note |
| 1.03 Recensione in rivista | 1.03 Review in journal |
| 1.04 Replica / breve intervento (e simili) | 1.04 Reply / brief note (and similar) |
| 1.05 Scheda bibliografica | 1.05 Bibliographic record |
| 1.06 Abstract in rivista | 1.06 Abstract in journal |
| 2.01 Capitolo / Saggio in libro | 2.01 Chapter / Essay in book |
| 2.02 Prefazione | 2.02 Preface |
| 2.03 Postfazione | 2.03 Afterword |
| 2.04 Breve introduzione | 2.04 Brief introduction |
| 2.05 Voce in dizionario o enciclopedia | 2.05 Entry in dictionary or encyclopedia |
| 2.06 Commento giuridico | 2.06 Legal commentary |
| 2.07 Scheda di catalogo | 2.07 Catalogue entry |
| 2.08 Recensione in volume | 2.08 Review in edited volume |
| 3.01 Monografia / trattato scientifico in forma di libro | 3.01 Monograph / Scientific treatise in book form |
| 3.02 Curatela | 3.02 Edited volume |
| 3.03 Edizione critica | 3.03 Critical edition |
| 3.04 Pubblicazione di fonti inedite | 3.04 Publication of unpublished sources |
| 3.05 Concordanze | 3.05 Concordances |
| 3.06 Indice | 3.06 Index |
| 3.07 Bibliografia | 3.07 Bibliography |
| 3.08 Edizione annotata/scolastica | 3.08 Annotated/educational edition |
| 4.01 Contributo in Atti di convegno | 4.01 Contribution in conference proceedings |
| 4.02 Riassunto (Abstract) | 4.02 Summary (Abstract) |
| 4.03 Poster | 4.03 Poster |
| 5.01 Libro (Traduzione) | 5.01 Book (Translation) |

| IRIS Type (Original – Italian) | IRIS Type (English Translation) |
|---|---|
| 5.02 Contributo in volume (Traduzione) | 5.02 Contribution in volume (Translation) |
| 5.03 Contributo in rivista (Traduzione) | 5.03 Contribution in journal (Translation) |
| 5.04 Traduzione di prodotti multimediali, teatrali, televisivi | 5.04 Translation of multimedia or theatrical products |
| 7.01 Carta tematica e geografica | 7.01 Thematic and geographic map |
| 7.02 Carta geologica | 7.02 Geological map |
| 7.03 Prodotto dell'ingegneria civile e dell'architettura | 7.03 Civil engineering and architecture product |
| 7.04 Software | 7.04 Software |
| 7.05 Banche dati | 7.05 Databases |
| 7.10 Prodotto artistico e spettacolare: Manufatto | 7.10 Artistic and performance product: Artifact |
| 7.11 Prodotto artistico e spettacolare: Prototipo d'arte e relativi progetti | 7.11 Artistic and performance product: Art prototype and related projects |
| 7.13 Rapporto tecnico | 7.13 Technical report |
| 7.14 Audiovisivi | 7.14 Audiovisual products |
| 7.15 Test psicologici | 7.15 Psychological tests |
| 8.03 Direzione di scavi archeologici | 8.03 Direction of archaeological excavations |